\documentclass[aps,pra,amsmath,preprint]{revtex4-1}

\usepackage{hyperref}
\usepackage{cleveref}
\usepackage{graphicx}
\usepackage{dcolumn}
\usepackage{bm}

\begin{document}

\title{Large Responsivity of Graphene Radiation Detectors With Thermoelectric Readout}

\author{A. Yurgens}
\email[]{yurgens-at-chalmers.se}
\affiliation{Department of Microtechnology and Nanoscience (MC2), Chalmers University of Technology, SE-412 96 G\"oteborg, Sweden}

\begin{abstract}
Simple estimations show that the thermoelectric readout in graphene radiation detectors can be extremely effective even for graphene with modest charge-carrier mobility $\sim 1000$~cm$^2$/(Vs).    The detector responsivity depends mostly on the residual charge-carrier density and split-gate spacing and can reach competitive values of $\sim 10^3 - 10^4$~V/W at room temperature. The optimum characteristics depend on a trade-off between the responsivity and the total device resistance.   Finding out the key parameters and their roles allows for simple detectors and their arrays, with high responsivity and sufficiently low resistance matching that of the radiation-receiving antenna structures.
\end{abstract}

\maketitle

\section{Introduction}
The graphene radiation detectors promise to be fast and sensitive devices in a broad frequency band from sub-THz- to infrared spectrum of electromagnetic radiation, operational from ambient~\cite{Vicarelli_2012}- to cryo temperatures \cite{Review_Koppens2014}. A negligibly small thermal mass of a typical graphene radiation absorber guarantees a very short response time of the detector \cite{Muller_resp-time_2011, TEP_2014, THz_mixing_hot-e, Lara_2019, GHz_mixer, Koppens_fs_nnano}. Several readout mechanisms in graphene detectors have been identified – bolometric \cite{quantum_dots}, thermoelectric (TEP) \cite{TEP_2014}, ballistic \cite{Teppe_ballistic}, based on noise thermometry \cite{noise_thermometry}, and electron-plasma waves \cite{resonant}, commonly called Dyakonov-Shur (D-S) mechanism \cite{Dyakonov_Shur, Dyakonov_Shur_PRL}.  However, resistivity of graphene changes significantly with temperature only in graphene samples with induced bandgap and only at low temperature. In the noise thermometry, the electronic temperature is obtained from first principles, but the measurement setups are complex and therefore impractical. Both ballistic and D-S mechanisms require very high mobility samples, in most cases obtained by laborious encapsulation of graphene in between hexagonal boron nitride (hBN) flakes. hBN of sufficiently high quality is unique and apparently available from only one laboratory in the World \cite{hBN}.

The TEP readout favorably stands out from the rest because of its simplicity, room-temperature operation, no electrical bias and therefore no 1/$f$ noise, scalable fabrication using CVD graphene, and undemanding electrical contacts. This combination of detector properties is particularly important for the fabrication of large detector arrays. The effectiveness of this readout stems from a high value of the Seebeck coefficient ($S\sim 100\ \mathrm{\mu V/K}$) \cite{TEP1,TEP2} and easy control over the charge-carrier density and sign in graphene.  An electrostatically induced \textit{p-n} junction gives an all-set access to the electronic temperature $T$ in graphene, thereby meeting the main requirement for radiation detectors. The temperature increase caused by incoming radiation is high because of a weak electron-phonon (e-ph) coupling in graphene \cite{Supercollision_cooling, e-cooling}. Combination of the weak e-ph coupling with large $S$ gives a strong foundation for building radiation-sensitive devices.

Among practical devices reported in the literature, graphene detectors with TEP readout experimentally demonstrated quite high responsivity $100-1000$~V/W and low noise-equivalent power $NEP \sim 20-200\ \mathrm{pW/\sqrt{Hz}}$ \cite{TEP_2014, Skoblin_SciRep, Skoblin_APL, GHz_mixer}. The responsivity $\Re$ in detectors with TEP readout is usually $10-100$ times higher than in those based on graphene field-effect transistors (GFET's), unless GFET's have very high mobility \cite{resonant}. The spread of device characteristics in the literature requires some qualitative understanding of key parameters that have the major effect on the detector performance. Here, estimations of limiting values of the TEP responsivity $\Re$ have been calculated by using earlier experimental data on electron cooling efficiency in graphene \cite{Betz_PhysRevLett, Supercollision_cooling}.   These estimations give basic guidelines on optimizing detectors with simple geometry and graphene of undemanding quality.
\begin{figure}[!ht]
\begin{center}
 \includegraphics[width=6cm]{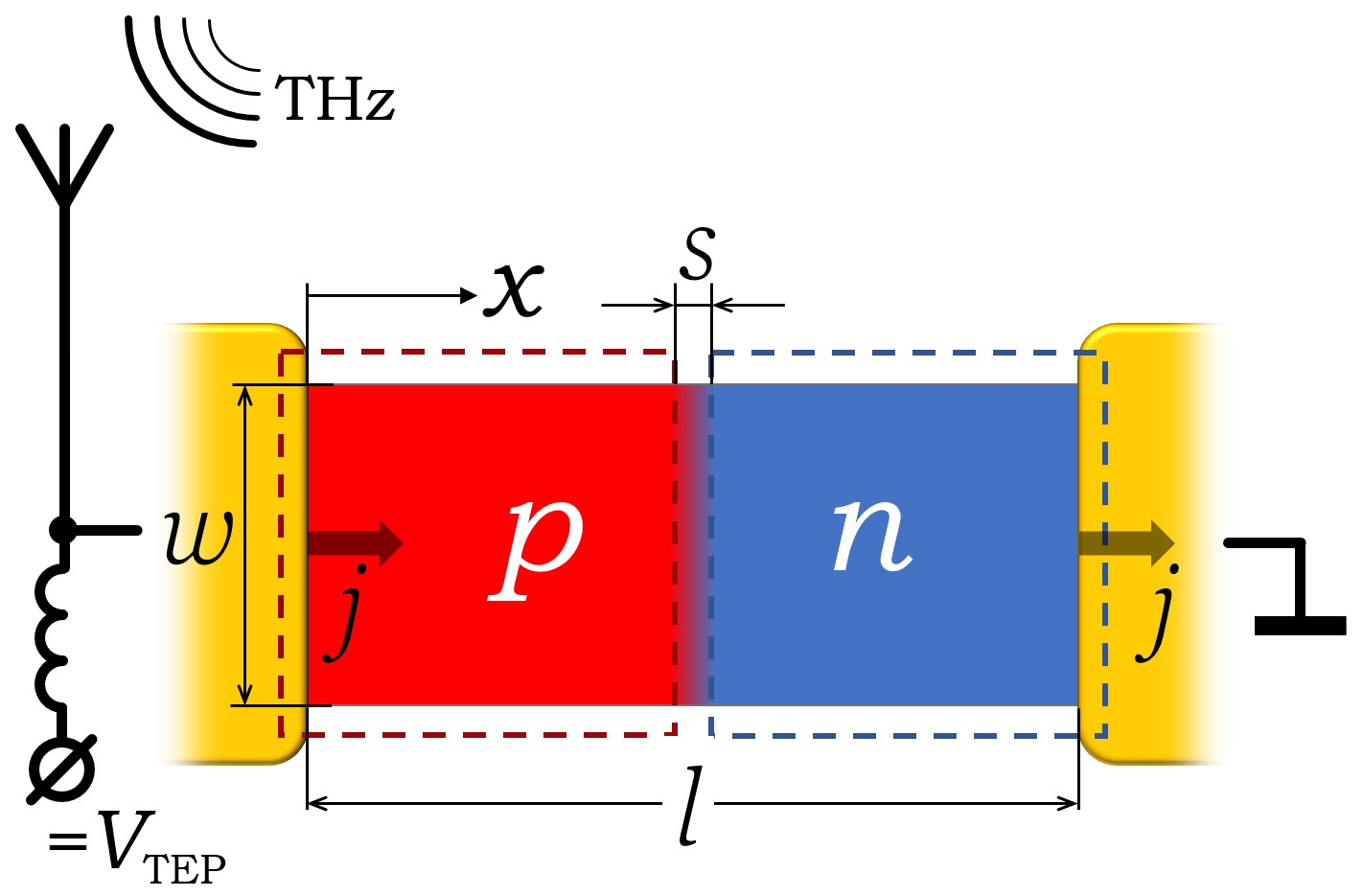}
 \vspace{-5mm}
\end{center}
  \caption{The model geometry of graphene detector with a \textit{p-n}-junction in the center. The $p$- and $n$ regions are induced electrostatically by a split gate with two parts (dashed rectangles) separated by the distance $s$. A current with linear density $j$ flows along the strip. }
  \label{fig:geometry}
\end{figure}

\section{Model}
The model geometry is shown in Fig.~\ref{fig:geometry}. A graphene strip of length $l$ and width $w$ is subdivided into  $p$- and $n$ regions. The strip rests on a substrate with an infinitely high thermal conductivity. The electrical current with the linear density $j$ flows in $x$ direction from the source- to drain electrodes made of thick metal films. The temperature $T_0$ of the substrate, electrodes, and phonons in graphene is assumed to be constant. The electrons in graphene are heated by the current and cooled by phonons through the electron-phonon interaction. The heating- and resulting temperature distribution $T(x)$ are highly non-uniform because of the spatially varying doping profile.

The charge-density (doping) profile $n_d(x)$ is approximated by:
\begin{equation}\label{eq:n-profile}
  n_d(x) = n_\mathrm{max}\Big[F\Big(\frac{x-l/2+s}{\delta }\Big) - F\Big(\frac{-x+l/2+s}{\delta }\Big)\Big],
\end{equation}
\noindent where $F(u)=1/[1+\exp(u)]$ is the Fermi function, $n_\mathrm{max}$ is the maximum induced doping, $s$ is the separation between the gates, and $\delta$ determines smearing of the profile due to fringing of the electric field at the gates edges; $\delta\sim$ the gate-dielectric thickness. In the conceivably possible case of chemical doping, $\delta$ would correspond to the lateral gradient of dopant concentration. For some brevity in equations, $n_d$ has sign, reflecting the sign of charge carriers in the $p$- and $n$ regions.

The region with the smallest $n_d$, i.e., the \textit{p-n} junction, has the lowest electrical conductance $\sigma$, which changes very little with temperature and therefore is taken to depend on $x$ only  (Eq.~\ref{eq:n-profile}):
\begin{equation}\label{eq:sigma}
\sigma (n_d) =\mu |e| \sqrt{n_d^2 + n_{0}^2},
\end{equation}
\noindent where $\mu$  is the charge-carrier mobility, $e$ is the elementary charge, and $n_0$ is the residual charge density,
\begin{equation}\label{eq:puddles}
n_0(T) = n_{00} + \beta T^2,
\end{equation}
\noindent where $n_{00}$ is the part resulting from the charge puddles~\cite{puddles} and the second term is due to smearing of the Fermi energy by temperature puddles\cite{puddles} and temperature with $\beta = (\pi/6) k_B^2/(\hbar v_F)^2$~\cite{intrinsic_n0}; $k_B$, $\hbar$, and $v_F$ are the Boltzmann- and Planck constants, and the Fermi velocity, respectively.  This model is described by the one-dimensional heat equation:
\begin{equation}
\label{eq:diffusion}
-\frac{\partial}{\partial x} \Big( \kappa_e \frac{\partial T}{\partial x}\Big) =
\frac{j^2}{\sigma} -
j T \frac{\partial S}{\partial x} -
\alpha_i \big( T^i-T_0^i \big),
\end{equation}
\noindent where $\kappa_e = L_0 \sigma T$ is the electronic sheet thermal conductivity and $L_0$ is the Wiedemann–Franz constant. The Seebeck coefficient $S$ in graphene is assumed to obey Mott's equation in the whole temperature range.
\begin{equation}
  S = \frac{\pi ^2}{3} \frac{k_B}{|e|} k_B T \frac{\partial\ln{\sigma }}{\partial n_d} \frac{\partial n_d}{\partial E_F}
\label{eq:Mott}
\end{equation}
\noindent where $k_B$ is the Boltzmann constant and $E_F$ is the Fermi energy.
The heat transfer to the phonon system is described by the last term in Eq.~\ref{eq:diffusion}. The exponent $i=3$ or $i=4$ at temperatures above or below the Bloch-Gr\"{u}neisen temperature $T_\mathrm{BG}$, respectively, and $\alpha_3 \propto n_d$ \cite{Supercollision_cooling}.

Numerically solving Eq.~\ref{eq:diffusion}  gives  $T(x)$,  the TEP voltage, and the total Joule dissipation for any bias current $j$. The current in real detectors is induced by the incoming radiation and is periodically varying with time $t$: $j(t)=j_0 \sin(\omega t)$. For $\omega\ll 2\pi/\tau$, where $\tau<50$~fs is the electron-heating time \cite{Koppens_fs_nnano}, the responsivity $\Re$ can be found by averaging the voltage and Joule power over one period of the ac bias:
\begin{eqnarray}
  V_\mathrm{TEP} &=& \Big\langle \int_0^l S(x)\nabla T(x) \mathrm{d}x \Big\rangle\\
  P_\mathrm{tot} &=& \Big\langle j^2 \int_0^l \frac{w\mathrm{d}x}{\sigma(x)} \Big\rangle\\
  \Re            &=& V_\mathrm{TEP}/P_\mathrm{tot}
\end{eqnarray}

\section{Results}
\begin{figure}[!htb]
\begin{center}
 \includegraphics[width=9cm]{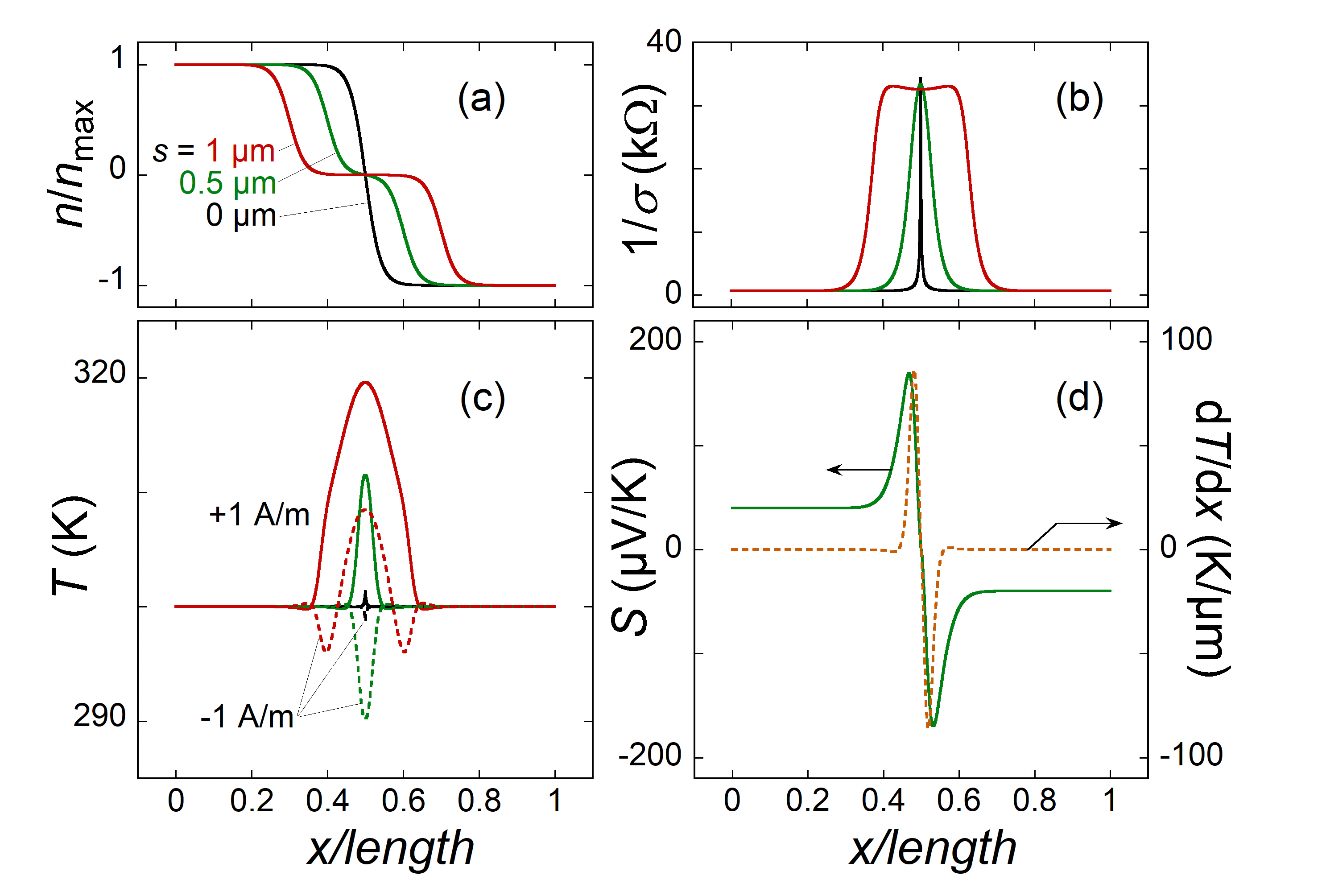}
 \vspace{-5mm}
\end{center}
  \caption{(a) The doping profile. (b) The local sheet resistance. (c) The temperature distributions for $j=+1$ and -1~A/m (solid and dashed curves, respectively). The Peltier effect is a significant source of the temperature variations. (d) The temperature gradient (orange) and Seebeck coefficient calculated from (a) and (c). Only narrow region around $x=l/2$ contributes to the output signal $V_\mathrm{TEP}$. The curves in (a)-(c) correspond to $s=0$, 0.5, and 1~$\mathrm{\mu m}$, (d)- to only $s=0.5\ \mathrm{\mu m}$; $n_{00}=10^{11}\ \mathrm{1/cm^2}$ for all panels; $j=+1$~A/m in (b) and (d).}
  \label{fig:overview}
\end{figure}

The following parameters were used in the calculations: $l\times w = 5\times 5\ \mathrm{\mu m^2}$, $\delta$ = 0.1 $\mathrm{\mu m}$, $s = 0...1\ \mathrm{\mu m}$ (see Fig.~\ref{fig:geometry}), $n_\mathrm{max}=1\times 10^{13}\ \mathrm{1/cm^2}$, $n_{00}=(1...100)\times 10^{10}\ \mathrm{1/cm^2}$,  $\mu=10^3$ or $10^4\ \mathrm{cm^2/(V s)}$, $T_0=$(4, 100, 200, 300~K), and $j_0=(0.01...1)$~A/m. For each $n_{00}$ and $T_0$, $j_0$ is chosen to restrict the maximum temperature rise at the \textit{p-n} junction to less than 10\% of $T_0$. Values of $T_0$ are picked in correspondence with the limiting cases of $T_0<T_\mathrm{BG}$ and $T_0>T_\mathrm{BG}$. The results of calculations for some combination of these parameters are shown in Fig.~\ref{fig:overview}.
Because of a low $\sigma$ at zero doping, the Joule heating is maximal in the center of the graphene strip. It is seen that $T(x)$ (Fig.~\ref{fig:overview}c) changes in agreement with the $n_d(x,s)$ curves (Fig.~\ref{fig:overview}a). The wider the region of zero doping the wider the $T(x)$.  The change from heating ($T(x)>T_0$) to cooling ($T(x)<T_0$) occurs because of the Peltier effect, which is a substantial source of temperature variation.  The temperature gradient and Seebeck coefficient are shown in (Fig.~\ref{fig:overview}d). The integral of their product gives the overall TEP signal. Clearly, only the parts of the strip where $dT/dx\neq 0$ contribute to the signal and it is favorable to have a smeared doping profile, i.e., larger $\delta$ and/or $s$ (see below).

 Fig.~\ref{fig:effect_n0} shows the effect of probably the most important parameter,  $n_{00}$, on $\Re$ at different $T_0$ and for $\mu=10^3$- and $10^4\ \mathrm{cm^2/(V s)}$. The responsivity significantly increases upon lowering  $n_{00}$, reaching a competitive value of $\geq 10^{4}\ \mathrm{V/W}$ for graphene with $\mu =10^3\ \mathrm{cm^2/(V s)}$. For a ten times higher $\mu$,  $\Re$ decreases roughly ten times. However, the advantage of having graphene with high mobility is about ten times lower overall resistance $R_\mathrm{tot}$  for $\mu=10^4$- than for $10^3\ \mathrm{cm^2/(V s)}$, about 1- and 10~k$\Omega$, respectively. This is important for impedance matching between graphene and a radiation-collecting antenna. Also, the temperature dependence of the residual charge density $n_0$ (Eq.~\ref{eq:puddles}) is significant. Without it, the responsivity gets unrealistically high $\Re >10^6$~V/W  (see the dash-dotted line in Fig.~\ref{fig:effect_n0}b). By dividing the thermal noise voltage variance per 1~Hz of bandwidth ($\sqrt{4 k_B T R_\mathrm{tot}}$) by the responsivity, the noise equivalent power ($NEP$) can be calculated.
\begin{figure}[!htb]
\begin{center}
 \includegraphics[width=8cm]{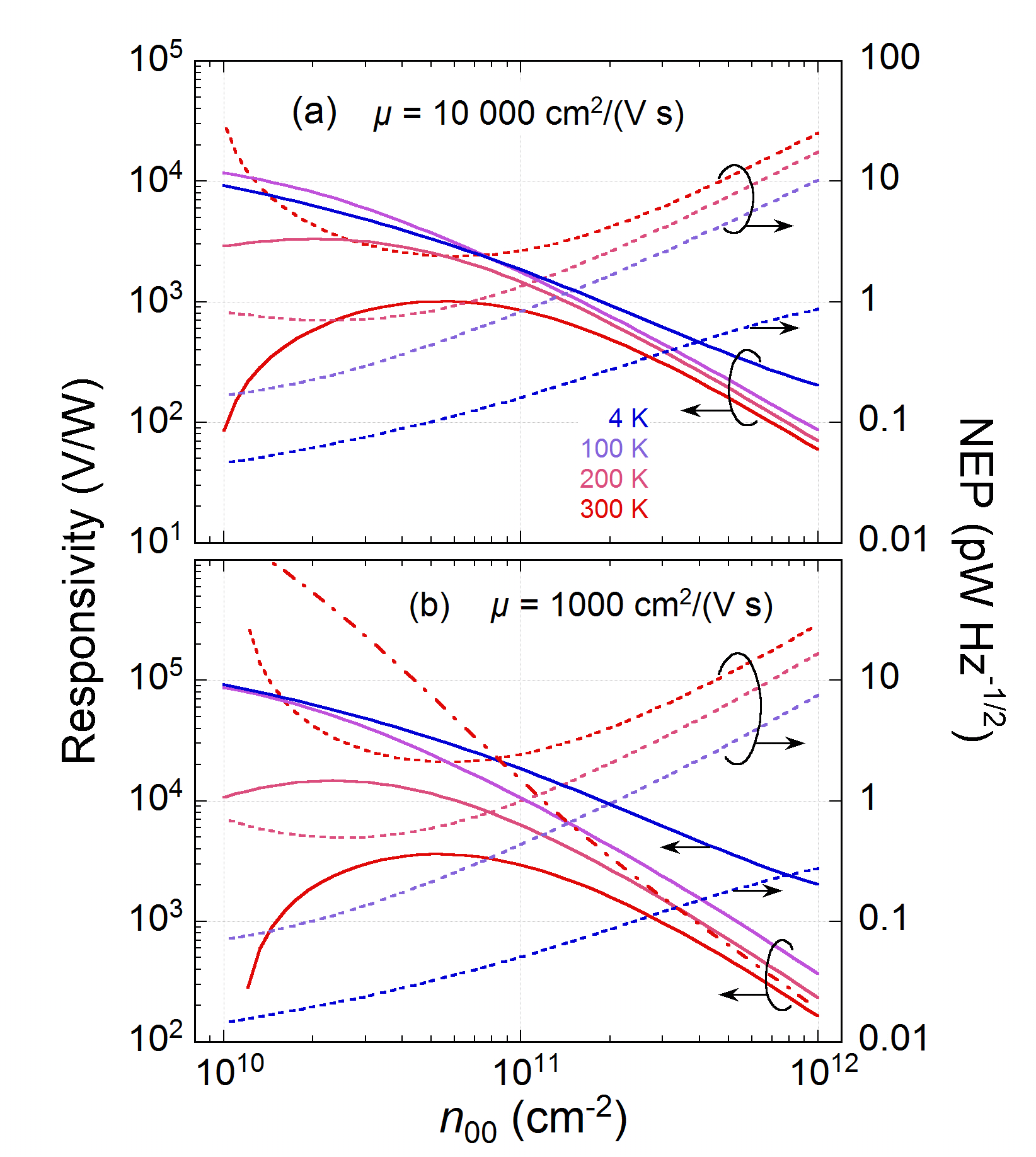}
 \vspace{-5mm}
\end{center}
  \caption{The responsivity $\Re$ (left ordinate, solid lines) and noise equivalent power $NEP$ (right, dotted lines) versus residual charge density due to charge puddles ($n_{00}$) for different ambient temperatures.  $\delta =0.1$, $s=0.5\ \mathrm{\mu m}$, $\mu=10^4$ (a) and $10^3\ \mathrm{cm^2/(V s)}$ (a). The curves for $T_0=4$~K have been calculated using $\alpha_4=0.5\ \mathrm{mW/(m^2 K^4)}$ (Ref.~\cite{Betz_PhysRevLett}) Ignoring the temperature dependence of $n_0$ (Eq.~\ref{eq:puddles}) gives too high $\Re>10^6$~V/W (dash-dotted line)} \label{fig:effect_n0}
\end{figure}

\begin{figure}[!htb]
\begin{center}
 \includegraphics[width=8cm]{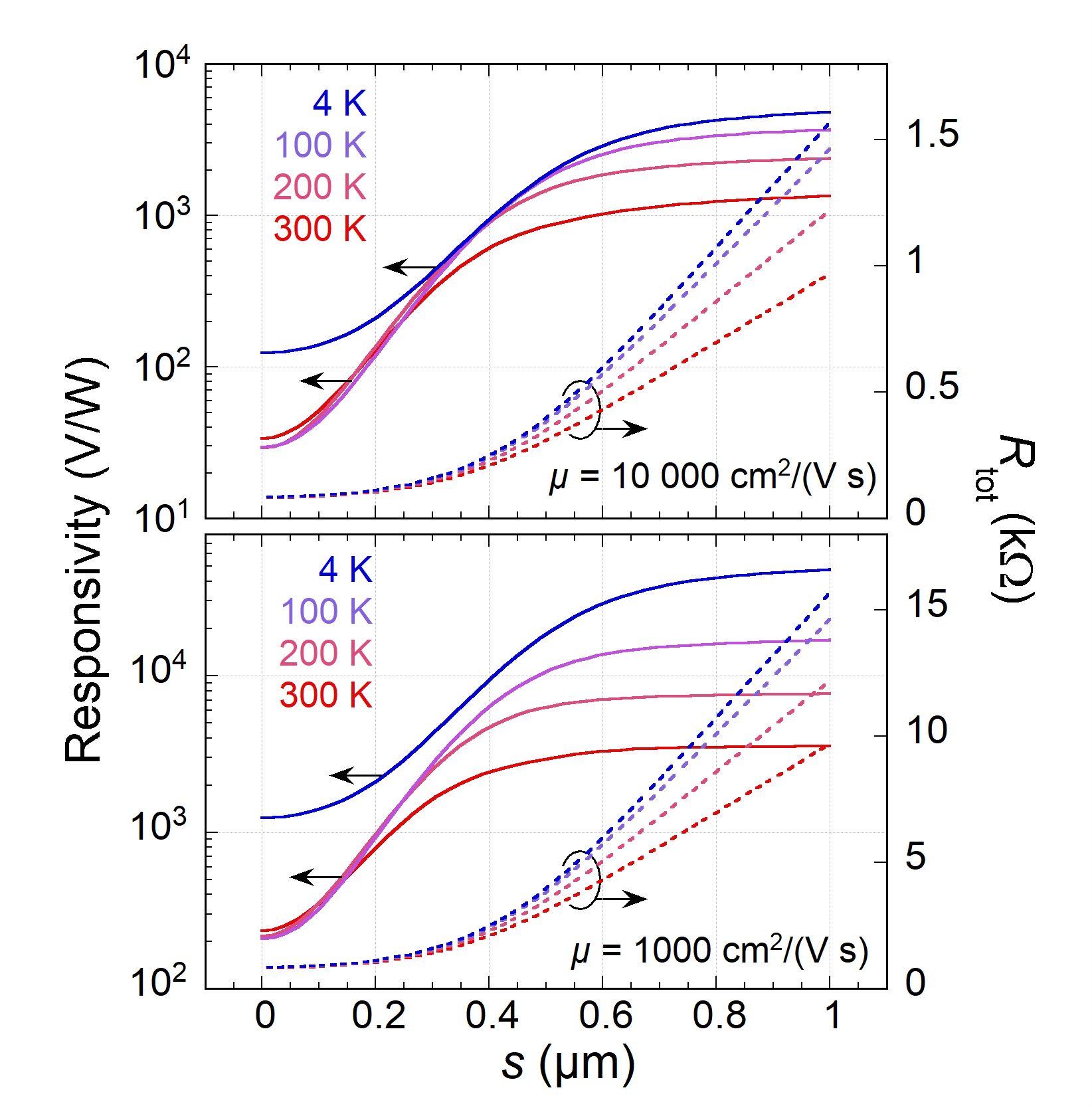}
 \vspace{-5mm}
\end{center}
  \caption{The responsivity $\Re$ (left ordinate) and overall resistance $R_{tot}$ (right) versus the split-gate separation  $s$ for $\delta=0.1\ \mathrm{\mu m}$, $\mu=10^4$ (top) and $10^3\ \mathrm{cm^2/(V s)}$ (bottom), and different $T_0 =4$, 100, 200, and 300~K.}
  \label{fig:effect_s}
\end{figure}

Next, the effect of split-gate separation $s$ is shown in Fig.~\ref{fig:effect_s}. The smearing of $T(x)$ increases with $s$ and is followed by a dramatic increase of $\Re$ at the expense of high $R_{\mathrm{tot}}$. At a relatively large $s$, the graphene channel is distinctly divided into three parts, \textit{p}, neutral, and \textit{n} (see Fig.~\ref{fig:overview}a), which is effectively equivalent to the \textit{p-n} junction extending in space. The increase of responsivity is then largely due to the increased Joule dissipation in the neutral region of graphene.

\section{Discussion}
As has been demonstrated above, $n_{00}$ is the main parameter governing $\Re$, which can be as high as $3\times 10^3$ V/W at $T=300$~K for $n_{00}\approx 5\times 10^{10}\ \mathrm{1/cm^2}$ and $\mu = 10^3\ \mathrm{cm^2/(V s)}$ (see Fig.~\ref{fig:effect_n0}). Of course, the majority of practical  devices have larger $n_{00}>10^{11}\ \mathrm{1/cm^2}$. Even then, $\Re\sim 10^3$, which is in agreement with experiments~\cite{Skoblin_APL}.
However, it has recently been found that the in graphene grown on SiC,  the residual doping can be very small, close to $n_0\sim 10^{10}\ \mathrm{1/cm^2}$ \cite{He2018}.  The temperature is then the main factor affecting $n_0$, resulting in a strong temperature dependence of graphene resistance $R(T)$, which in turn allows for the development of a low-temperature bolometer mixer \cite{Lara_2019}. Note, that if the TEP readout were used instead of the bolometric one, a high $\Re\sim 10^5$~V/W could possibly be reached at much higher $T_0\sim 100$~K (see Fig.~\ref{fig:effect_n0}).

\begin{figure}[!htb]
\begin{center}
 \includegraphics[width=7cm]{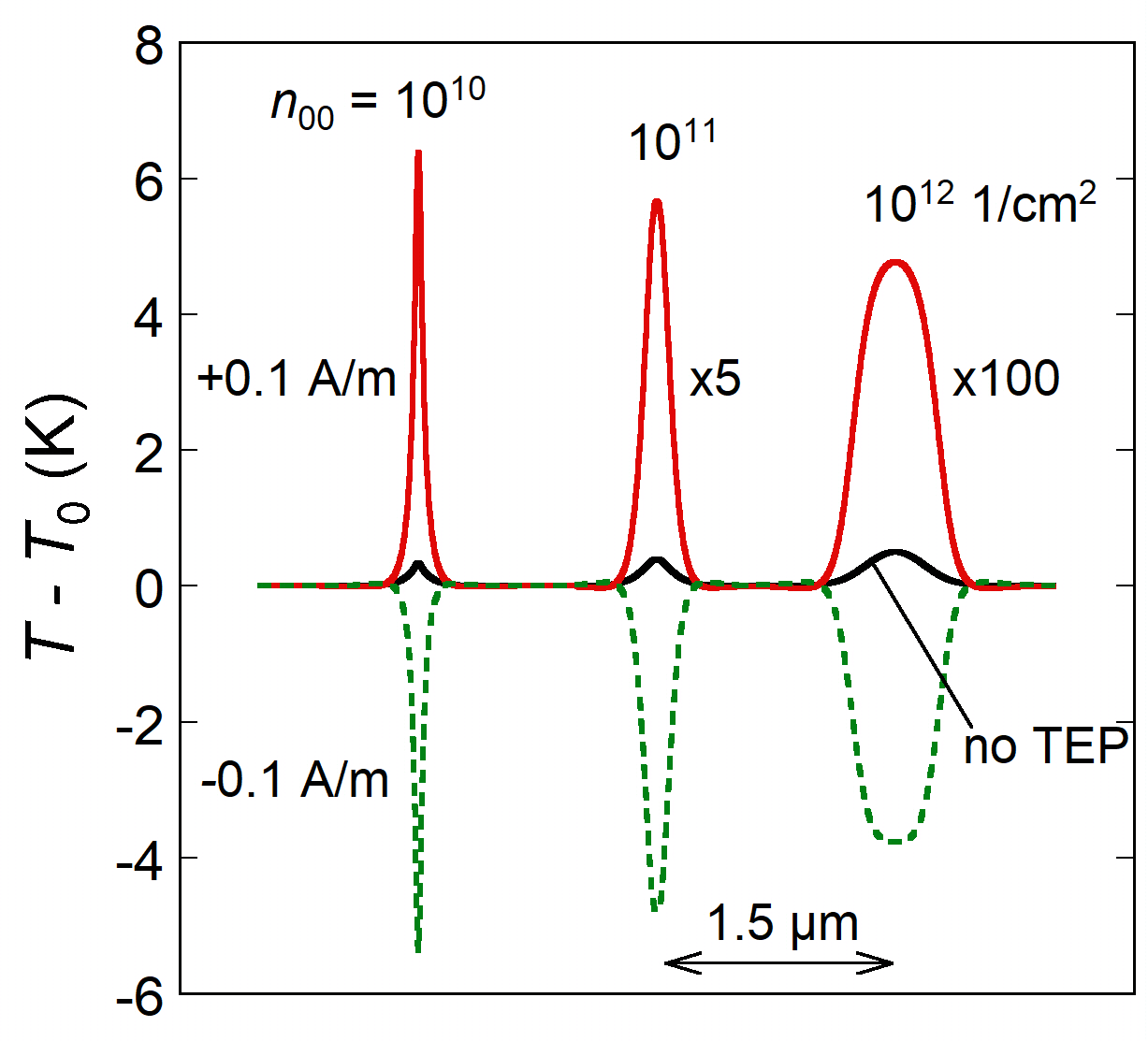}
  \vspace{-5mm}
\end{center}
  \caption{The temperature change $T(x)-T_0$ corresponding to heating (red) and cooling (green) of the \textit{p-n} junction by $\pm0.1\mathrm{-A/m}$ current for $n_{00}=10^{10}, 10^{11},\  \mathrm{and}\ 10^{12}\ \mathrm{1/cm^2}$ (left to right). The curves are shifted horizontally for clarity.  $\mu =1000\ \mathrm{cm^2/(V s)}$, $\delta = 0.1\ \mathrm{\mu m}$, $s=0.5\ \mathrm{\mu m}$, and $T_0=100$~K. The black curves correspond to Joule heating only, without the Peltier effect.  }
  \label{fig:effect_Peltier}
\end{figure}

The Peltier effect is obviously dominant in heating of graphene \textit{p-n} junctions, especially for sufficiently uniform graphene with $n_{00}\leq 10^{11}\ \mathrm{1/cm^2}$. Fig.~\ref{fig:effect_Peltier} shows the temperature distribution for different $n_{00}$. For comparison, there are curves corresponding to the thermoelectric effects switched off. It is noteworthy that the temperature in a \textit{p-n} junction would change much more dramatically than if only Joule heating were considered.  This is also clear from the comparison between the first (Joule) and the second (Peltier) terms in Eq.~\ref{eq:diffusion}. The latter is typically 10-100 times larger than the former at high $T_0$.  This prompts for using \textit{p-n} junctions in graphene as thermal sources of infrared light \cite{radiation_control}.

The TEP readout, because of its open-circuit condition, is expected to be limited by the thermal Johnson–Nyquist noise only, contrary to other types of readout using some bias current. In the presence of bias current, the 1/$f$ noise starts to dominate. The 1/$f$ noise in graphene is rather high and extends to frequencies $\sim 10^5$~Hz \cite{1-f_noise}, which can be a problem for detector systems with mechanical beam choppers. Fig.~\ref{fig:effect_n0} shows that the $NEP$ does not change much with increasing $\mu$ - the reduced responsivity is compensated by a lower thermal noise because of a smaller $R_\mathrm{tot}$ at high $\mu$. For a typical $n_{00} \sim 2-4 \times 10^{11}$~1/cm$^2$, $NEP$ lies between 1 and 10 pW/Hz$^{1/2}$ at high temperature. This is at least ten times better than the $NEP's$ of other types of uncooled direct detectors~\cite{Sizov_2018}. These estimations are also in agreement with the recent experimental works on TEP readout, e.g.~\cite{TEP_2014,Skoblin_SciRep,Skoblin_APL}.

Even in GFET-based detectors, the TEP readout mechanism can contribute with a substantial, if not dominating, signal.   The detection signal in GFET's is proportional to the transfer characteristic of GFET, i.e., to $d\ln(\sigma) /dV_g$, with $V_g$ being the gate voltage. The same combination of values is involved in Mott's equation (Eq.~\ref{eq:Mott}), given that $n_d \propto V_g$. This makes these detection mechanisms difficult to tell apart and/or totally exclude the TEP contribution to the signal.
Indeed, the top gate subdivides the graphene channel into three regions with conceivably different doping: \textit{p-p'-p} or \textit{n-n'-n}. If graphene globally is close to the charge neutrality point, the doping of different regions can have opposite signs, \textit{p-n-p} or \textit{n-p-n}, thereby creating two \textit{p-n} junctions in series. When the AC current, which is injected via the gate, flows predominantly towards either the source or drain, only one of the \textit{p-n} junctions will be heated by the current. This breaks the symmetry and gives rise to an uncompensated TEP signal. The farther apart the \textit{p-n} and \textit{n-p} junctions (i.e., the wider the gate) the more asymmetric is the current through the junctions and the higher the responsivity can be expected. Compare for instance Refs.~\cite{Vicarelli_2012} and \cite{NL2014_Stake}, with the gate widths ($\Re$) of 0.3~$\mu$m (0.1~V/W) and 2.5~$\mu$m (14~V/W), respectively.   No wonder that $\Re$ in the latter case was larger even though a CVD graphene with $\mu <2000\ \mathrm{cm^2/(V s)}$ was used \cite{NL2014_Stake}. For graphene with very high mobility \cite{resonant}, it is however tempting to speculate that the TEP- and D-S detection mechanisms might become mixed together, possibly resulting in responsivity amplification. This however requires a thorough theoretical analysis.

The present work assumed that the current was applied through the metal electrodes while ignoring their contact resistances to graphene. The contact resistance does not seem to be a big problem for the low-ohmic edge contacts, which are possible for graphene encapsulated in both hBN \cite{contacts_Wang2013} and Parylene \cite{Skoblin_SciRep, Skoblin_Parylene}, the latter being important for scaling up the device fabrication. Also, the contact resistance can be effectively reduced by increasing the perimeter of contact even for a non-encapsulated graphene \cite{contacts_Lemme2019}. Finally, the capacitive coupling of antennas to graphene, where the contact resistance does not play any role for the open-circuit TEP readout, has recently been realized \cite{Skoblin_APL}.

\section{Conclusion}
The effectiveness of the thermoelectric readout mechanism in graphene radiation detectors has been estimated for a few key parameters, assuming a simple device geometry.  The residual charge density and sharpness of the \textit{p-n} junction are the main parameters that affect the detector performance most. In all cases, there is a trade-off between the responsivity and the total device resistance.  Concluding, the thermoelectric readout in graphene radiation detectors represents a very competitive platform for building simple and sensitive direct detectors of radiation and arrays of them.

\section*{Acknowledgment}
This work was supported by the FLAG-ERA grant DeMeGRaS and the Korea-Sweden research programme on flexible arrays of graphene radiation detectors with thermoelectric readout.

\end{document}